\documentclass[jou,natbib,floatsintext]{apa7}
\usepackage[utf8]{inputenc}

\usepackage{url,graphicx,siunitx,booktabs,comment,multirow,amsmath,subcaption,rotating,longtable,lscape,xcolor}
\usepackage{authblk}
\DeclareCaptionTextFormat{tabletext}{\hspace{-\parindent}\textit{#1}}

\newcommand{\spcellc}[2][c]{%
	\begin{tabular}[#1]{@{}c@{}}#2\end{tabular}}

\title{Automated Evaluation Of Psychotherapy Skills\\ Using Speech And Language Technologies}
\shorttitle{Automated Evaluation Of Psychotherapy Skills Using Speech And Language Technologies}
\leftheader{Flemotomos et al.}

\authorsnames[1,2,1,2,2,1,3,4,5,1,6,7,8,3,7,{1,2,4}]{Nikolaos Flemotomos,Victor R. Martinez,Zhuohao Chen,Karan Singla,Victor Ardulov,Raghuveer Peri,Derek D. Caperton,James Gibson,Michael J. Tanana,Panayiotis Georgiou,Jake Van Epps,Sarah P. Lord,Tad Hirsch,Zac E. Imel,David C. Atkins,Shrikanth Narayanan}
\authorsaffiliations{
	{Department of Electrical and Computer Engineering, University of Southern California, Los Angeles, CA, USA},
	{Department of Computer Science, University of Southern California, Los Angeles, CA, USA},
	{Department of Educational Psychology, University of Utah, Salt Lake Ciity, UT, USA},
	{Behavioral Signal Technologies Inc., Los Angeles, CA, USA},
	{College of Social Work, University of Utah, Salt Lake Ciity, UT, USA},
	{University Counseling Center, University of Utah, Salt Lake Ciity, UT, USA},
	{Department of Psychiatry and Behavioral Sciences, University of Washington, Seattle, WA, USA},
	{Department of Art + Design, Northeastern University, Boston, MA, USA}}

\authornote{
James Gibson performed this work while at the University of Southern California.

David C. Atkins, Shrikanth Narayanan, Zac E. Imel, and  Panayiotis Georgiou conceived the idea of the described system. James Gibson developed a first prototype of the processing pipeline. Nikolaos Flemotomos, Victor R. Martinez, Zhuohao Chen, Karan Singla, Victor Ardulov, and Raghuveer Peri worked on training, adapting, and evaluating the various individual modules and the end to end system. Michael J. Tanana helped testing the system in real-world settings, which led to several revisions and improvements. Jake Van Epps played a crucial role on data collection. Derek D. Caperton and Sarah P. Lord led the efforts in human behavioral coding. Tad Hirsch led the efforts of designing the graphical interface. Nikolaos Flemotomos reviewed the literature and drafted the original manuscript. All authors provided critical input, discussion, and revision of the manuscript.

Michael J. Tanana, Tad Hirsch, Zac E. Imel, David C. Atkins, and Shrikanth Narayanan are co-founders with equity stake in a technology company, Lyssn.io, focused on tools to support training, supervision, and quality assurance of psychotherapy and counseling. Sarah P. Lord also holds an equity stake in Lyssn.io. Shrikanth Narayanan is Chief Scientist and co-founder with equity stake in Behavioral Signal Technologies, a company focused on creating technologies for emotional and behavioral machine intelligence. James Gibson is a Senior Machine Learning Engineer in Behavioral Signal Technologies. The remaining authors report no conflicts of interest.

Correspondence concerning this article should be addressed to Nikolaos Flemotomos, Department of Electrical and Computer Engineering, University of Southern California, Los Angeles, CA 90089. 
Contact: flemotom@usc.edu}

\abstract{With the growing prevalence of psychological interventions, it is vital to have measures which rate the effectiveness of psychological care to assist in training, supervision, and quality assurance of services. Traditionally, quality assessment is addressed by human raters who evaluate recorded sessions along specific dimensions, often codified through constructs relevant to the approach and domain. This is however a cost-prohibitive and time-consuming method that leads to poor feasibility and limited use in real-world settings. To facilitate this process, we have developed an automated competency rating tool able to process the raw recorded audio of a session, analyzing who spoke when, what they said, and how the health professional used language to provide therapy.  Focusing on a use case of a specific type of psychotherapy called Motivational Interviewing, our system gives comprehensive feedback to the therapist, including information about the dynamics of the session (e.g., therapist’s vs. client’s talking time), low-level psychological language descriptors (e.g., type of questions asked), as well as other high-level behavioral constructs (e.g., the extent to which the therapist understands the clients’ perspective). We describe our platform and its performance using a dataset of more than 5,000 recordings drawn from its deployment in a real-world clinical setting used to assist training of new therapists. Widespread use of automated psychotherapy rating tools may augment experts' capabilities by providing an avenue for more effective training and skill improvement, eventually leading to more positive clinical outcomes.
}


\begin{document}
\maketitle
\section*{Need for Psychotherapy Quality Assessment Tools}
Recent epidemiological research suggests that developing a mental disorder is the norm, rather than the exception, estimating that the lifetime prevalence of diagnosable mental disorders (i.e., the proportion of the population that, at some point in their life, have experienced or will experience a mental disorder) is around $50\%$~\citep{kessler2005lifetime} or even more~\citep{schaefer2017enduring}. According to data from 2018, an estimated 47.6 million adults in the United States had some mental illness, while only about 1 in 7 adults received mental health services~\citep{substance2019}. 
	
Psychotherapy is a commonly used process in which mental health disorders are treated through communication between an individual and a trained mental health professional. Even though the positive effects of psychotherapy have been well documented~\citep{lambert2002effectiveness,weisz1995effects,perry1999effectiveness}, there is room for improvement in terms of the quality of services provided. In particular, a substantial number of patients report negative outcomes, with signs of mental health deterioration after the end of therapy~\citep{klatte2018adverse,curran2019how}. Apart from patient characteristics~\citep{lambert2002effectiveness}, therapist factors play a significant and clinically important role in contributing to negative outcomes~\citep{saxon2017contribution}. This has direct implications for more rigorous training and supervision~\citep{lambert1997effectiveness}, quality improvement, and skill development. A critical factor that can lead to increased performance and thus ensure high quality of services is the provision of accurate feedback to the practitioner~\citep{hattie2007power}. This can take various forms; both client progress monitoring~\citep{lambert2018collecting} and performance-based feedback ~\citep{schwalbe2014sustaining} have been reported to reduce therapeutic skill erosion and to contribute to improved clinical outcomes. The timing of the feedback is of utmost importance as well, since it has been shown that immediate feedback is more effective than delayed~\citep{kulik1988timing}.

In psychotherapy practice, however, providing regular and immediate performance evaluation is almost impossible. Behavioral coding---the process of listening to audio recordings and/or reading session transcripts in order to observe therapists' behaviors and skills~\citep{bakeman2012behavioral}---is both time-consuming and cost-prohibitive when applied in real-world settings. It has been reported~\citep{moyers2005assessing} that, after intensive training and supervision that lasts on average 3 months, a proficient coder would need up to two hours to code just a $20\si{min}$-long session of Motivational Interviewing (MI), a specific type of psychotherapy which is also the focus of the current study. The labor-intensive nature of coding means that the vast majority of psychotherapy sessions are not evaluated. As a result, many providers get inadequate feedback on their therapy skills after their initial training~\citep{miller2006disseminating} and behavioral coding is mainly applied for research purposes with limited outreach to community settings~\citep{proctor2011outcomes}. At the same time, the barriers imposed by manual coding usually lead to research studies with relatively small sample sizes~\citep{magill2014technical}, limiting progress in the field. It is, thus, made apparent that being able to evaluate a therapy session and provide feedback to the practitioner at a low cost and in a timely manner would both boost psychotherapy research and scale up quality assessment to real-world use. In the current work, we investigate whether it is feasible to analyze a therapy session recording in a fully automatic way and provide rich feedback to the therapist within short time.

\section*{Behavioral Coding for Motivational Interviewing}
Motivational Interviewing (MI)~\citep{miller2012motivational}, often used for treating addiction and other conditions, is a client-centered intervention that aims to help clients make behavioral changes through resolution of ambivalence. It is a psychotherapy treatment with evidence supporting that specific skills are correlated with the clinical outcome~\citep{gaume2009counselor,magill2014technical} and also that those skills cannot be maintained without ongoing feedback~\citep{schwalbe2014sustaining}. Thus, great effort from MI researchers has been devoted to developing instruments to evaluate fidelity to MI techniques.

The gold standard for monitoring clinician fidelity to treatment is behavioral observation and coding~\citep{bakeman2012behavioral}. During that process, trained coders assign specific labels or numeric values to the psychotherapy session, which are expected to provide important therapy-related details (e.g., ``how many open questions were posed by the therapist?'' or ``did the counselor accept and respect the client's ideas?'') and essentially reflect particular therapeutic skills. While there are a variety of coding schemes~\citep{madson2006measures}, in this study we focus on a widely used research tool, the Motivational Interviewing Skill Code (MISC 2.5;~\citealp{houck2010misc}), which was specifically developed for use with recorded MI sessions~\citep{madson2006measures}. MISC defines behavior codes both for the counselor and the patient, but for the automated system reported in this paper we focus on counselor behaviors.

The MISC manual~\citep{houck2010misc} defines both session-level and utterance-level codes. The session-level (or ``global'') codes characterize the entire interaction and are scored on a 5-point Likert scale ranging from 1 (\textit{poor}) to 5 (\textit{excellent}). Table~\ref{table:globalMISC_descr} gives an overview of the six therapist-related global MISC ratings with a short description for each one. When coding at the utterance-level, instead of assigning numerical values, the coder decides in which behavior category each utterance belongs. An utterance is a ``thought unit''~\citep{houck2010misc}, which means that multiple consecutive phrases might be parsed into a single utterance and, likewise, multiple utterances might compose a single sentence or talk turn. After the session is parsed into utterances, each one is assigned one of the codes summarized in Table~\ref{table:utteranceMISC_descr} (or gets the label NC if it can not be coded).

\begin{table*}[ht]
	\caption{Therapist-related session-level codes, as defined by MISC 2.5. Each code is scored on a 5-point Likert scale.\\\hspace*{5cm}}\vspace*{-.4cm}
	\label{table:globalMISC_descr}
	\centering
	\begin{tabular}{ll}
		\toprule
		name & high score means that counselor... \\
		\midrule
		acceptance & consistently communicates acceptance and respect to the client\\
		empathy & makes an effort to accurately understand the client’s perspective\\
		direction & is focused on a specific target behavior\\
		autonomy support & does not attempt to control the client’s behavior or choices\\
		collaboration & interacts with their clients as partners and avoids an authoritarian attitude \\
		evocation & tries to ``draw out'' client's own desire for changing\\
		\bottomrule
	\end{tabular}    
\end{table*}

	\begin{table*}[ht]
		\centering
		\caption{Therapist-related utterance-level codes, as defined by MISC 2.5. Most of the examples are drawn from the MISC manual~\citep{houck2010misc}. Many of the code assignments depend on the client's previous utterance (C).}
		\label{table:utteranceMISC_descr}
		\begin{tabular}{lll}
			\toprule
			abbr. & name & example \\
			\midrule
			ADP & Advise with Permission & Would it be all right if I suggested something?\\
			ADW & Advise w/o Permission & I recommend that you attend 90 meetings in 90 days.\\
			AF & Affirm & Thank you for coming today.\\
			CO & Confront & (C: I don’t feel like I can do this.) Sure you can.\\
			DI & Direct & Get out there and find a job.\\
			EC & Emphasize Control & It is totally up to you whether you quit or cut down.\\
			FA & Facilitate & Uh huh. (\textit{keep-going acknowledgment})\\
			FI & Filler & Nice weather today!\\
			GI & Giving Information & Your blood pressure was elevated [...] this morning.\\
			QUO & Open Question & Tell me about your family.\\
			QUC & Closed Question & How often did you go to that bar?\\
			RCP & Raise Concern with Permission & Could I tell you what concerns me about your plan?\\
			RCW & Raise Concern w/o Permission & That doesn’t seem like the safest plan.\\
			RES & Simple Reflection & (C: The court sent me here.) That’s why you’re here.\\
			REC & Complex Reflection & (C: The court sent me here.) This wasn't your choice to be here.\\
			RF & Reframe & (C: [...] something else comes up [...]) You have clear priorities.\\
			SU & Support & I'm sorry you feel this way.\\
			ST & Structure & Now I’d like to switch gears and talk about exercise.\\
			WA & Warn & Not showing up for court will send you back to jail.\\
			\midrule
			NC & No Code & You know, I… (\textit{meaning is not clear})\\
			\bottomrule
		\end{tabular}    
	\end{table*}

The platform we present is evaluated under real-world conditions, by continuously gathering and analyzing psychotherapy sessions recorded in the counseling center of an American university with a large student body.  Our system is part of a broader study where the goal is to investigate whether therapists make more extensive use of MI techniques after MI-related training and we thus evaluate all the recorded sessions following the MISC protocol.

\section*{Psychotherapy Evaluation in the Digital Era}
Psychotherapy sessions are interventions primarily based on spoken language, which means that the information capturing the session quality is encoded in the speech signal and the language patterns of the interaction. Thus, with the rapid technological advancements in the fields of Speech and Natural Language Processing (NLP) over the last few years (e.g.,~\citealp{xiong2017toward,devlin2019bert}), and despite many open challenges specific to the healthcare domain~\citep{quiroz2019challenges}, it is not surprising to see trends in applying computational techniques to automatically analyze and evaluate psychotherapy sessions.
	
Such efforts span a wide range of psychotherapeutic approaches including couples therapy~\citep{black2013toward}, MI~\citep{xiao2016behavioral} and cognitive behavioral therapy~\citep{flemotomos2018language}, used to treat a variety of conditions such as addiction~\citep{xiao2016behavioral} and post-traumatic stress disorder~\citep{shiner2012automated}. Both text-based~\citep{xiao2012analyzing,imel2015computational} and audio-based~\citep{black2013toward,xiao2014modeling} behavioral descriptors have been explored in the literature and have been used either unimodally or in combination with each other~\citep{singla2018using}.

In this study we focus on behavior code prediction from textual data. Most research studies focused on text-based behavioral coding have relied on written text excerpts~\citep{barahona2018deep} or used manually-derived transcriptions of the therapy session~\citep{lee2019identifying,can2015dialog,gibson2019multi}. However, a fully automated evaluation system for deployment in real-world settings requires a speech processing pipeline that can analyze the audio recording and provide a reliable speaker-segmented transcript of what was spoken by whom. This is a necessary condition before such an approach is introduced into clinical settings since, otherwise, it may eliminate the burden of manual behavioral coding, but it introduces the burden of manual transcription. Transcription errors introduced by Automatic Speech Recognition (ASR) algorithms may have a significant effect on the performance of NLP-based models~\citep{malik2018performance}, so demonstrating the practical feasibility of a fully automated pipeline is an important task.

An end-to-end system is presented by~\cite{xiao2015rate} and~\cite{xiao2016technology}, where the authors report a case study of automatically predicting the empathy expressed by the provider. A similar platform, focused on couples therapy, is presented by~\cite{georgiou2011thats}. Even employing an ASR module with relatively high error rate, those systems were reported to provide competitive prediction performance~\citep{georgiou2011thats}. The scope of the particular studies, though, was limited only to session-level codes, while the evaluation sessions were selected from the two extremes of the coding scale. Thus, for each code the problem was formulated as a binary classification task trying to identify therapy sessions where a particular code (or its absence) is represented more prominently (e.g., identify `low' vs. `high' empathy).

\subsection*{Current Study}
\begin{figure*}[ht]
	\begin{subfigure}{\textwidth}
		\centering
		\includegraphics[width=.65\textwidth]{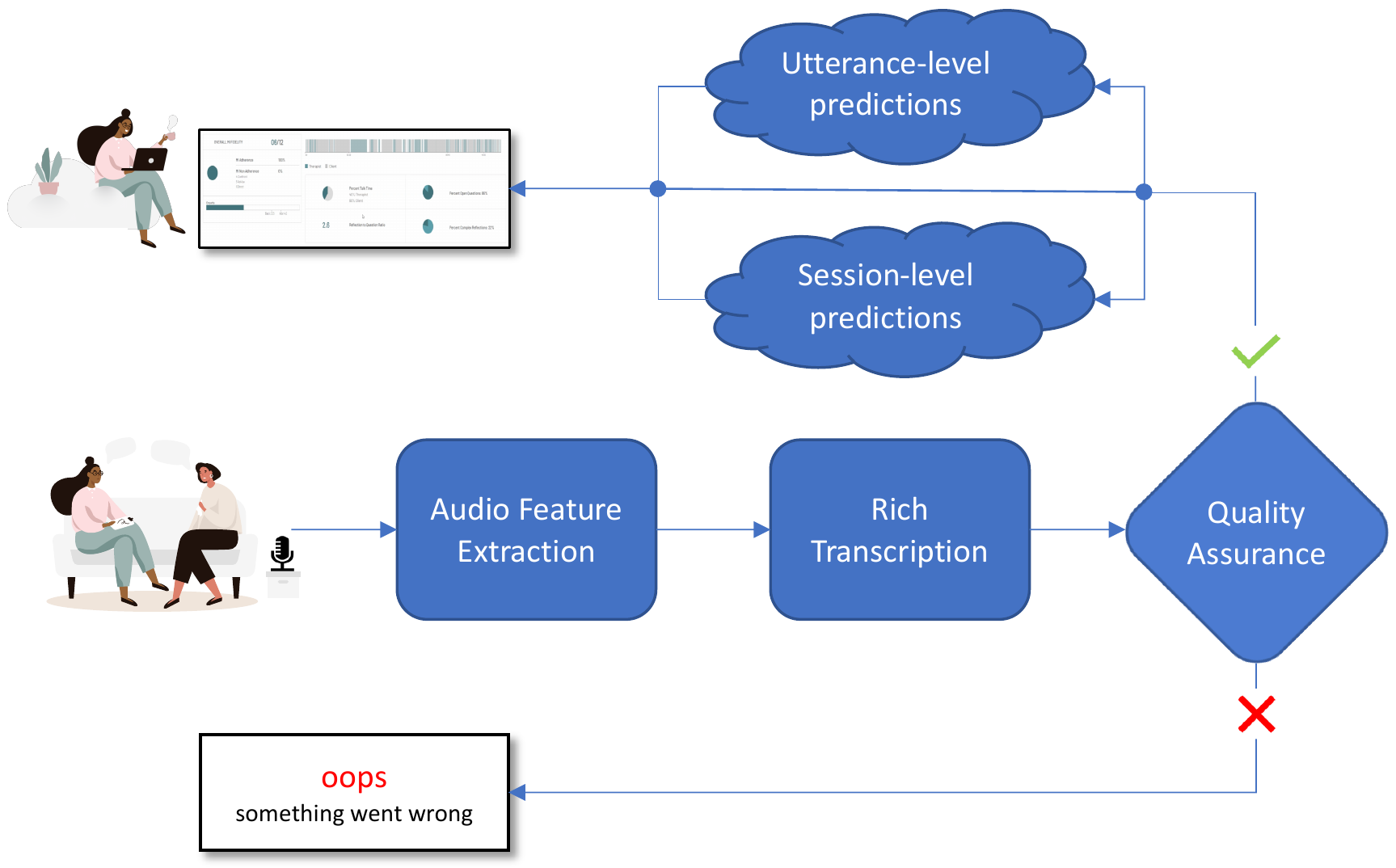}
		\caption{}
		\label{fig:overall}
	\end{subfigure}
	\begin{subfigure}{\textwidth}
		\centering
		\includegraphics[width=.93\textwidth]{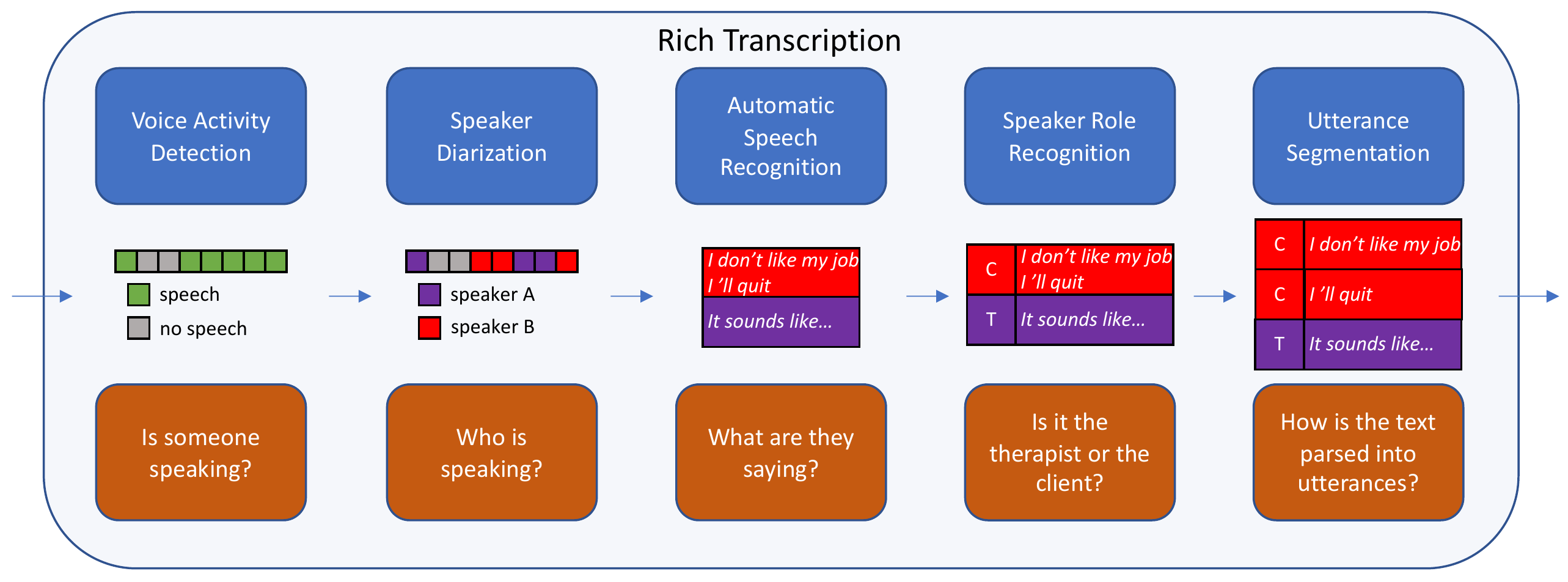}
		\caption{}
		\label{fig:rich_trans}
	\end{subfigure}
	\caption{(a) Overview of the system used to assess the quality of a psychotherapy session and provide feedback to the therapist. Once the audio is recorded, it is automatically transcribed to find who spoke when and what they said. If the transcription meets certain quality criteria, this textual information is used to predict utterance-level and session-level behavior codes which are summarized into an interactive feedback report. Otherwise, an error message is displayed to the user.\\
		(b) Rich transcription module. The dyadic interaction is transcribed through a pipeline that extracts the linguistic information encoded in the speech signal and assigns each speaker turn to either the therapist or the client.}
	\label{fig:overview}
\end{figure*}

In the current paper we demonstrate and analyze a platform (Figure~\ref{fig:overview}) able to process a raw recording of a psychotherapy session and provide, within short time, performance-based feedback according to therapeutic skills and behaviors expressed both at the utterance and at the session level. We focus on dyadic psychotherapy interactions (i.e., one therapist and one client) and the quality assessment is based on the counselor-related codes of the MISC protocol~\citep{houck2010misc}. The behavioral codes are predicted by NLP algorithms that analyze the linguistic information captured in the automatically derived transcriptions of the session. 

The overall architecture is illustrated in Figure~\ref{fig:overall}. After both parties have formally consented, the therapist begins recording the session. The digital recording is directly sent to the processing pipeline and appropriate acoustic features are extracted from the raw speech signal. The rich audio transcription component of the system (Figure~\ref{fig:rich_trans}) consists of five main steps: (a)~Voice Activity Detection (VAD), where speech segments are detected over silence or background noise, (b)~speaker diarization, where the speech segments are clustered into same-speaker groups (e.g.,  speaker A, speaker B of a dyad), (c)~Automatic Speech Recognition (ASR), where the audio speech signal of each speaker-homogeneous segment is transcribed to words, (d)~Speaker Role Recognition (SRR), where each speaker group is assigned their role: in our case study, ‘therapist’ or ‘client’, and (e)~utterance segmentation, where the speaker turns are parsed into utterances which are the basic units of behavioral coding. The generated transcription is used to estimate a variety of behavior codes both at the utterance and at the session level, which reflect target constructs related to therapist behaviors and skills.

The behavioral analysis of the counselor is summarized into a comprehensive feedback report provided through an interactive web-based platform~\citep{hirsch2018s,imel2019design}. Through the platform, the user is able to review the raw MISC predictions of the system (e.g., empathy score and utterances labeled as reflections), several theory-driven functionals of those (e.g., ratio of questions to reflections), session statistics (e.g., ratio of client's to therapist's talking time), as well as the entire speaker-segmented transcription, accompanied by the corresponding audio recording. Additionally, the user is given the option to take notes and make comments linked to specific timestamps or utterances. That way, the platform can be used directly by the provider as a self-assessment method or by a supervisor as a supportive tool that helps them deliver more effective and engaging training. 

Since the system was designed with real-world deployment in mind, it was important to incorporate specific confidence metrics which reflect the quality of the automatic transcription.Employing quality safeguards helps us both identify potential computational errors, and determine whether the input was an actual therapy session or not (e.g., whether the therapist pushed the recording button by mistake). If certain quality thresholds are not met, then the final report is not generated and feedback is not provided for the specific session. Instead, an error message is displayed to the counselor. For example, in a scenario where speaker segmentation fails because the recording is too noisy or the two speakers have very similar acoustic characteristics, the system would not know which utterances correspond to the provider and which correspond to the client; as a result, the subsequent prediction algorithms would fail to accurately capture counselor-related behaviors. Being able to avoid such scenarios is of crucial importance for a system used in clinical settings. 

As illustrated in Figure~\ref{fig:overview}, we have chosen a pipelined implementation of the system, as opposed to a more convoluted architecture, potentially able to predict behavioral codes directly from the speech waveform. That way, we are able to provide a feedback report containing much richer information than merely the behavior codes or statistics of those. In particular, the user has access to the entire transcript and can understand how particular behaviors are linked to the linguistic content of the corresponding utterances. This design increases the interpretability and, as a result, the trust of the clinical provider to the system. Additionally, we are able to extract and provide information critical for the quality assessment of the therapy session, not directly related to behavior codes, such as the client's speaking time. Finally, the quality assurance of the generated transcription is based on certain quality safeguards (described later in the paper) corresponding to specific sub-modules of the pipeline, such as the VAD and the diarization. So, if a potential error is detected at an early stage of the pipeline (e.g., VAD), the entire processing can be halted, thus avoiding wasting computational resources.

\section*{{Materials and Methods}}
\subsection*{Datasets}
The design of the system presented in this work is based on datasets drawn from a variety of sources. We have combined large speech and language corpora both from the psychotherapy domain and from other fields (meetings, telephone conversations, etc.). That way, we wanted to ensure high in-domain accuracy when analyzing psychotherapy data, but also robustness across various recording conditions. In order to use and evaluate the system in real-world clinical settings, we have additionally collected and analyzed a set of more than 5,000 recordings of therapy sessions between a provider and a patient at a University Counseling Center (UCC). The details of the various datasets are presented in the following sections.

\subsubsection*{Out-of-Domain Corpora}
\paragraph*{Audio Sources}
The acoustic modeling performed in this work was mainly based on a large collection of speech corpora, widely used by the research community for a variety of speech processing tasks. Specifically, we used the Fisher English~\citep{cieri2004fisher}, ICSI Meeting Speech~\citep{janin2003icsi}, WSJ~\citep{paul1992design}, and 1997 HUB4~\citep{graff19971996} corpora, available  through the Linguistic Data Consortium (LDC), as well as Librispeech~\citep{panayotov2015librispeech}, TED-LIUM~\citep{rousseau2014enhancing}, and AMI~\citep{carletta2005ami}. This combined speech dataset consists of more than 2,000h of audio and contains recordings from a variety of scenarios, including business meetings, broadcast news, telephone conversations, and audiobooks/articles. 

\paragraph*{Text Sources}
The aforementioned datasets are accompanied by manually-derived transcriptions which can be used for language modeling tasks. In our case, since we need to capture linguistic patterns specific to the psychotherapy domain, the main reason we need some out-of-domain text corpus is to build a background model that guarantees a large enough vocabulary and minimizes the unseen words during evaluation. To that end, we use the transcriptions of the Fisher English corpus, featuring a vocabulary of 58.6K words and totaling more than 21M tokens. 

\subsubsection*{Psychotherapy-Related Corpora}
\paragraph*{Audio Sources}
n order to train and adapt our machine learning models, used both for the transcription component of the system and for the behavior coding predictions, we also used several psychotherapy-focused corpora. In particular, we used a collection of 337 MI sessions (for which audio, transcription and manual coding information were available) from six independent clinical trials (ARC,  ESPSB, ESP21, iCHAMP, HMCBI, CTT). In more detail, ARC (9 sessions;~\citealp{tollison2008questions}), ESPSB (38 sessions;~\citealp{lee2014randomized}) and ESP21 (19 sessions;~\citealp{neighbors2012randomized}) feature brief alcohol interventions. CTT (194 sessions;~\citealp{baer2009agency}) also consists of alcohol interventions, but using standardized patients (i.e., actors portraying patients). Finally, iCHAMP (7 sessions;~\citealp{lee2013indicated}) addresses marijuana addiction and HMCBI (70 sessions;~\citealp{krupski2012testing}) addresses poly-drug abuse. We refer to the combined dataset as the TOPICS-CTT corpus and we have split it into train ($\text{TOPICS-CTT}_{train}$; 242 sessions) and test ($\text{TOPICS-CTT}_{test}$; 95 sessions) sets.

The mean duration of the sessions is $29.10\si{min}$ (std=$15.65\si{min}$).  The number of unique therapists and clients recorded in those sessions is given in Table~\ref{table:ther_cl_topics}. Unfortunately, the client IDs are not available for the HMCBI sessions, so the exact total number of different clients is not known. However, under the assumption that it is highly improbable for the same client to visit different therapists in the same study, and having the necessary metadata available for the rest of the corpus, we make the train/test split in a way that we are highly confident there is no overlap between	speakers. This is important since we want to make sure that our models capture universal behavior-specific patterns during training and not speaker-specific linguistic information.

\begin{table*}[ht]
	\centering
	\caption{{Number of sessions, unique therapists, and unique clients in the 6 clinical trials composing the TOPICS-CTT corpus. The client IDs are not known for the HMCBI data.}}
	\label{table:ther_cl_topics}
	\begin{tabular}{lcccccc}
		\toprule
		&ARC & ESPSB & ESP21 & iCHAMP & HMCBI & CTT\\
		\midrule
		\#sessions& 9 & 38 & 19 & 7 & 70 & 194\\
		\#therapists& 3 & 15 & 8 & 5 & 15 & 132\\
		\#clients& 9 & 38 & 19 & 7 & -- & 4 \\
		\bottomrule 
	\end{tabular}
\end{table*}

\paragraph*{Text Sources}
The transcripts of the aforementioned MI sessions were enhanced by data provided by the Counseling and Psychotherapy Transcripts Series (CPTS), available from the Alexander Street Press (\url{alexanderstreet.com}) via library subscription. This included transcripts from a variety of therapy interventions totaling about 300K utterances and 6.5M words. For this corpus, no audio or behavioral coding are available, and the data were hence used only for language-based modeling tasks.

\subsubsection*{University Counseling Center Data Collection}
Through a collaboration with the university-based counseling center of a large western university, we gathered a corpus of real-world psychotherapy sessions to evaluate the proposed system. Therapy treatment was provided by a combination of licensed staff as well as trainees pursuing clinical degrees. Topics discussed span a wide range of concerns common among students, including depression, anxiety, substance use, and relationship concerns. All the participants (both patients and therapists) had formally consented to their sessions being recorded. Study procedures were approved by the institutional review board of the University of Utah. Each session was recorded by two microphones hung from the ceiling of the clinic offices, one omni-directional and one directed to where the therapist generally sits. 

Data reported in this article were collected between September, 2017 and March, 2020, for a total of 5,097 recordings. Every time a session is recorded, it is automatically sent to the audio processing pipeline, and a performance-based feedback report is generated.
We note that some of those recordings were not actually valid therapy sessions (e.g., the therapist pushed the recording button by mistake); however we do have relevant safeguards for such cases, as described later in the article. Eventually, 4,268 sessions were successfully processed with a mean duration of $49.77\si{min}$ (std=$11.50\si{min}$), giving a therapy corpus totaling more than 2.8M utterances and 28M words (according to the automatically generated output), including sessions from at least 59 therapists and 1040 clients (there are a few sessions for which such metadata are not available).

In order to adapt and evaluate the pipeline, 188 sessions were selected to be manually transcribed and coded. The coding took place in two independent trials (one in mid 2018 and one in late 2019), with some differences in the procedure between the two. For the first coding trial (96 sessions), the transcriptions were stripped of punctuation and coders were asked to parse the session into utterances. During the second trial (92 sessions), the human transcriber was asked to insert punctuation, which was used to assist parsing. Additionally, for the second batch of transcriptions, stacked behavioral codes (more than one code per utterance) were allowed in case one of the codes is question (QUC or QUO). Because of those differences in the coding approach, we are reporting results independently for the two trials; in particular, we have split the first trial into train ($\text{UCC}_{train}$; 50 sessions), development ($\text{UCC}_{dev}$; 26 sessions), and test ($\text{UCC}_{test_1}$; 20 sessions) sets, while we refer to the second trial as the $\text{UCC}_{test_2}$ set and we only use it for evaluation. That way, we are able to monitor the robustness of the system through time, without continuously adapting to new data. For similar reasons as in the case of the TOPICS-CTT corpus, the split for the first trial was done in a way so that there is no speaker overlap between the different sets.

Each of the 188 sessions was coded by at least one of three coders. Among those, 14 sessions (from the first trial) were coded by two or three coders, so that we can have a measure of inter-rater reliability (IRR). To that end, we estimated Krippendorff’s alpha ($\alpha$)~\citep{krippendorff2018content} for each code, a statistic which is generalizable to different types of variables and flexible with missing observations~\citep{hallgren2012computing}. Since sessions were parsed into utterances from the human raters, the unit of coding is not fixed, so we got an estimate for the utterance-level codes at the session level by using the per session occurrences of each label. For the IRR analysis, we treated the occurrences of the utterance-level codes as ratio variables and the values of the session-level codes as ordinal variables. The results for all the codes are given in Table~\ref{table:interrater_alpha}. For the session-level codes, the `within one’ reliability is also provided, since it is recommended that only a distance between the raters’ different scores greater than one point in the Likert scale should be considered disagreement~\citep{schmidt2019lessons}.

\begin{table*}[ht]
	\centering
	\caption{Krippendorff’s alpha ($\alpha$) to estimate inter-rater reliability for the utterance-level (upper 4 tables; ratio measurement level) and the session-level (lower table; ordinal measurement level) codes in UCC data. For the utterance-level codes, we get an estimate through their per-session occurrences. For the session-level codes, the `within one' agreement is also provided, demonstrating whether the distance between the raters' different scores was at most one point in the Likert scale. $^*$ denotes that the particular code was not used (count=0) by at least 2 coders for at least half of the analyzed sessions. RCP was never used by any coder.}
	\label{table:interrater_alpha}
	\begin{tabular}{ll}
		\toprule
		code & IRR ($\alpha$) \\
		\midrule
		ADP & 0.542$^*$\\
		ADW & 0.422\\
		AF & 0.123\\
		CO & 0.497$^*$\\
		DI & 0.590\\
		\bottomrule
	\end{tabular}
	\begin{tabular}{ll}
		\toprule
		code & IRR ($\alpha$) \\
		\midrule
		EC & 0.558\\
		FA & 0.868\\
		FI & 0.784\\
		GI & 0.861\\
		QUO & 0.945\\
		\bottomrule
	\end{tabular}
	\begin{tabular}{ll}
		\toprule
		code & IRR ($\alpha$) \\
		\midrule
		QUC & 0.897\\
		RCP & --$^*$\\
		RCW & 0.000$^*$\\
		RES & 0.268\\
		REC & 0.478\\
		\bottomrule
	\end{tabular}
	\begin{tabular}{ll}
		\toprule
		code & IRR ($\alpha$) \\
		\midrule
		RF & 0.093$^*$\\
		SU & 0.345\\
		ST & 0.434\\
		WA & -0.054$^*$\\
		& \\
		\bottomrule
	\end{tabular}\\
	\begin{tabular}{lcc}
		\toprule
		code & IRR ($\alpha$) & IRR `within 1' ($\alpha$) \\
		\midrule
		acceptance & 0.468 & 0.747\\
		empathy & 0.532 & 1.000\\
		direction & 0.593 & 0.795\\
		aut. support & 0.464 & 0.743\\
		collaboration & 0.287 & 0.472\\
		evocation & 0.410 & 0.626\\
		\bottomrule
	\end{tabular}
\end{table*}

\subsection*{Data pre-processing}
The manually transcribed UCC sessions do not contain any timing information, which means that we needed to align the provided audio with text. That way, we were able to get estimates of the ``ground truth'' information required to evaluate some of the modules of our system, such as VAD and diarization. We did so by using the Gentle forced aligner (\url{github.com/lowerquality/gentle}), an open-source, Kaldi-based~\citep{povey2011kaldi} tool, in order to align at the word level. However, we should note that this inevitably introduces some error to the evaluation process, since $9.4\%$ of the words per session on average (std=$3.4\%$) remain unaligned.

Another pre-processing step we needed to take in order to have a meaningful evaluation of the system on the UCC data is related to the behavioral labels assigned by the humans and by the platform. In particular, some of the utterance-level MISC codes are assigned very few times within a session by the human raters and the corresponding IRR is very low (Table~\ref{table:interrater_alpha}); additionally, there are pairs or groups of codes with very close semantic interpretation as reflected by the examples in Table~\ref{table:utteranceMISC_descr} (e.g., REC and RF). Thus, we clustered the codes into composite groups resulting in 9 target labels. The mapping between the codes defined in the MISC manual and the target labels, as well as the occurrences of those labels in the UCC data, is given in Table~\ref{table:misc_human_pc_map}. Comparing Tables~\ref{table:interrater_alpha} and~\ref{table:misc_human_pc_map}, we see that IRR is substantially higher, on average, after this grouping. The code FA seems to dominate the data, because most of the verbal fillers (e.g., uh-huh, mm-hmm, etc.)---which are very frequent constructs in conversational speech---and single-word utterances (e.g., yeah, right, etc.) are labeled as FA.

\begin{table*}[ht]
		\centering
		\caption{Mapping between the MISC-defined behavior codes and the grouped target labels, together with the occurrences of each group in the training and development UCC sets. The inter-rater reliability for the grouped labels is also given in terms of the Krippendorff’s alpha ($\alpha$) value.}
		\label{table:misc_human_pc_map}
		\begin{tabular}{ccccc}
			\toprule
			group & MISC codes & count ($\text{UCC}_{train}$) & count ($\text{UCC}_{dev}$) & IRR ($\alpha$)\\
			\midrule
			FA & FA & 5581 & 2500 & 0.868\\
			GI & GI, FI & 3797 & 1695 & 0.898\\
			QUC & QUC & 1911 & 693 & 0.897\\
			QUO & QUO & 1116 & 405 & 0.946\\
			REC & REC, RF & 2212 & 1041 & 0.479\\
			RES & RES & 609 & 155 & 0.268\\
			MIN & ADP, ADW, CO, DI, RCW, RCP, WA & 479 & 163 & 0.606\\
			MIA & AF, EC, SU & 428 & 238 & 0.363\\
			ST & ST & 542 & 135 & 0.434\\
			\bottomrule
		\end{tabular}    
\end{table*}

It should be noted that, even though we are only dealing with individual therapy sessions between a provider and a client, sometimes more than two speakers appear in a recording (e.g., a third speaker interrupts the session). However, for our analyses, we never took this piece of information into consideration (i.e., for diarization and speaker role recognition we always assume two speakers).

\subsection*{Audio Feature Extraction}
For all the modules of the speech pipeline (VAD, diarization, ASR), the acoustic representation is based on the widely used Mel Frequency Cepstrum Coefficients (MFCCs). For the UCC data, the channels from the two recording microphones are combined through acoustic beamforming~\citep{anguera2007acoustic}, using the open-source BeamformIt tool (\texttt{github.com/xanguera/BeamformIt}). 

\subsection*{Automatic Rich Transcription}
Before proceeding to the automatic behavioral coding, we need to transcribe the raw audio recording, in order to get information about the content, the speakers, and the utterance boundaries. This is not just a pre-processing step allowing us to apply NLP algorithms, but it also provides invaluable information which will be later incorporated in the final feedback report (e.g., talking time of each speaker). The rich transcription pipeline we propose is illustrated in Figure~\ref{fig:rich_trans}. In the following sections, we describe the various sub-modules of the system. Further technical details are provided in Appendix~\ref{app:tech_detals}.

\subsubsection*{Voice Activity Detection}
The first step of the transcription pipeline is to extract the voiced segments of the input audio session. The rest of the session is considered to be silence, music, background noise, etc., and is not taken into account for the subsequent steps. To that end, a two-layer feed-forward neural network is used giving a frame-level probability. This is a pre-trained model, initially developed as part of the SAIL lab efforts for the Robust Automatic Transcription of Speech (RATS) program~\citep{thomas2015improvements}. The model was trained to reliably detect speech activity in highly noisy acoustic scenarios, with most of the noise types included during training being military noises like machine gun, helicopter, etc. Hence, in order to make the model better suited to our task, the original model was adapted using the $\text{UCC}_{dev}$ data. Optimization of the various parameters was done with respect to the unweighted average recall. The frame-level outputs are smoothed via a median filter and converted to longer speech segments which are passed to the diarization sub-system. During this process, if the silence between any two contiguous voiced segments is less than 0.5sec, the corresponding segments are merged together. 

\subsubsection*{Speaker Diarization}
Speaker diarization answers the question ``who spoke when'' and it traditionally consists of two steps. First, the speech signal is partitioned into segments where a single speaker is present. Then, those speaker-homogeneous segments are clustered into same-speaker groups. For this work we follow the x-vector/PLDA paradigm, an approach known to achieve state-of-the-art performance for speaker recognition and diarization~\citep{sell2018diarization,snyder2018x}. In particular, each voiced segment, as predicted by VAD, is partitioned uniformly into subsegments and for each subsegment a fixed-dimensional speaker embedding (x-vector) is extracted. Once the x-vectors have been extracted, an affinity matrix is constructed with the pairwise distances between the subsegments. The similarity metric used is based on the Probabilistic Linear Discriminant Analysis (PLDA) framework~\citep{ioffe2006probabilistic,prince2007probabilistic}, within which each data point is considered to be the output of a model which incorporates both within-individual and between-individual variation. The subsegments are finally clustered together according to Hierarchical Agglomerative Clustering (HAC). The assumption here is that each session has exactly two speakers (i.e. therapist vs. client), so we continue the HAC procedure until two clusters have been constructed. As a post-processing step after diarization, adjacent speech segments assigned to the same speaker are concatenated together into a single speaker turn, allowing a maximum of $1\si{sec}$ in-turn silence. 

\subsubsection*{Automatic Speech Recognition}
After we get the speaker-homogeneous segments from the diarization module, we need to extract the linguistic content captured within each segment, since this will be the information supplied to the subsequent text-based algorithms. ASR depends on two components; the Acoustic Model (AM), which calculates the likelihood of acoustic observations given a sequence of words, and the Language Model (LM), which calculates the likelihood of a word sequence by describing the distribution of typical language usage. 

In order to train the AM, we build a Time-Delay Neural Network (TDNN) with subsampling~\citep{peddinti2015time}, an architecture which has been successfully applied in conversational speech achieving remarkable performance~\citep{peddinti2015jhu}. To increase robustness against different speakers, the input features are augmented by i-vectors~\citep{saon2013speaker}, extracted online through a sliding window. The network is trained on a large combined speech dataset composed of the Fisher English, ICSI  Meeting  Speech, WSJ, 1997 HUB4, Librispeech, TED-LIUM, AMI, and TOPICS-CTT corpora. Among those, TED-LIUM and the clean portion of Librispeech are augmented with speed perturbation, noise, and reverberation~\citep{ko2015audio}. The final combined, augmented corpus contains more than 4,000 hours of phonetically rich speech data, recorded under different conditions and reflecting a variety of acoustic environments. The ASR AM was built and trained using the Kaldi speech recognition toolkit~\citep{povey2011kaldi}. 

In order to build the LM, we independently train two 3-gram models using the SRILM toolkit~\citep{stolcke2002srilm}. One is trained with in-domain psychotherapy data from the CPTS transcribed sessions. This is interpolated with a large background model, in order to minimize the unseen words during the sytem employment. The background LM is trained with the Fisher English corpus, which features conversational telephone data.

\subsubsection*{Speaker Role Recognition}
After diarization has been performed, we have the entire set of utterances clustered into two groups; however, there is not a natural correspondence between the cluster labels and the actual speaker roles (i.e., therapist and client). For our purposes, Speaker Role Recognition (SRR) is exactly the task of finding the mapping between the two. Even though different speaker roles follow distinct patterns across various modalities (e.g., audio, language, structure), the linguistic stream of information is often the most useful for the task in hand~\citep{flemotomos2018combined}. So, in this work we are focusing on this modality, provided by the ASR output.   

Let's denote the two clusters which have been identified by diarization as $S_1$ and $S_2$, each one containing the utterances assigned to the two different speakers. We know a priori that one of those speakers is the therapist (T) and one is the client (C). In order to do the role matching, two trained LMs, one for the therapist (LM$_T$) and one for the client (LM$_C$), are used. We then estimate the perplexities of $S_1$ and $S_2$ with respect to the two LMs and we assign to $S_i$ the role that yields the minimum perplexity. In case one role minimizes the perplexity for both speakers, we first assign the speaker for whom we are most confident. The confidence metric is based on the absolute distance between the two estimated perplexities~\citep{flemotomos2018language}. The required LMs are 3-gram models trained with the SRILM toolkit~\citep{stolcke2002srilm}, using the TOPICS-CTT$_{train}$ and CPTS corpora.

\subsubsection*{Utterance Segmentation}
The output of the ASR and SRR modules is at the segment level, with the segments defined by the VAD and diarization algorithms. However, silence and speaker changes are not always the right cues to help us distinguish between utterances, which are the basic units of behavioral coding. The presence of multiple utterances per speaker turn is a challenge we often face when dealing with conversational interactions. Especially in the psychotherapy domain, it has been shown that the right utterance-level segmentation can significantly improve the performance of automatic behavior code prediction~\citep{chen2020feature}.

Thus, we have included an utterance segmentation module at the end of the automatic transcription, before employing the subsequent NLP algorithms. In particular, we merge together all the adjacent segments belonging to the same speaker in order to form speaker-homogeneous talk-turns, and we then segment each turn using the DeepSegment tool (\url{github.com/notAI-tech/deepsegment}). DeepSegment has been designed to perform sentence boundary detection having specifically ASR outputs in mind, where punctuation is not readily available. In this framework, sentence segmentation is viewed as a sequence labeling problem, where each word is tagged as being either at the beginning of a sentence (utterance), or anywhere else. DeepSegment addresses the problem employing a Bidirectional Long-Short Term Memory (BiLSTM) network with a Conditional Random Field (CRF) inference layer~\citep{ma2016end}.

\subsection*{Quality Assurance}
The goal of the current study is to provide accurate and reliable feedback to the counselor in a real-world environment. Thus, it is essential that we ensure we do not produce feedback reports which are problematic, either because of bad audio quality, or because of errors during computation. We have identified that most of the errors are produced during the first steps of the processing pipeline and are propagated to the subsequent steps. Thus, we have incorporated simple quality safeguards, able to catch errors associated with the audio recording, the VAD, or the diarization. Specifically, before any further processing, the following conditions need to be met:
\begin{enumerate}
    \item The duration of the entire recording has to be between $60\si{sec}$ and $5\si{h}$. Given that a typical therapy session in our study is about $50\si{min}$-long, a session outside this range indicates either that the provider pushed the recording button by mistake, or that they forgot to stop recording.
    \item At least $25\%$ of the session has to be flagged as voiced, according to the VAD output. During a typical conversational interaction, there are pauses of silence which are especially useful in psychotherapy~\citep{levitt2001sounds}. Although silence is an essential aspect of communication, the distribution of the silence gaps' duration is highly skewed with most of them being very short~\citep{heldner2010pauses}. If most of the therapy session is flagged as unvoiced, this is an indication of bad audio quality, of some inherent error of the VAD algorithm employed, or of a prolonged audio file where the therapist forgot to stop the recording after the actual session.
    \item The average duration of the voiced segments cannot be longer than 20sec. Even though words are the primary means of communication, silence gaps are not just useful, but necessary in order for spoken language to be meaningful and natural. When our VAD system is incapable of detecting unvoiced segments, it is usually an indication of bad audio quality.
    \item The minimum percentage of speech assigned to each speaker is $10\%$ of the total speaking time. Since we are dealing with dyadic conversational scenarios, it is expected that each of the two speakers talks for a substantial amount of time. Even though therapy is not a normal dialog and the provider often plays more the role of the listener~\citep{hill2009helping}, if a person seems to be talking for less than $10\%$ of the time (e.g., less than about $5\si{min}$ in a typical $50\si{min}$-long session), then we are highly confident there is some problem. This may be an issue associated either with the audio quality, or with high speaker error introduced by the diarization module because the two speakers have similar acoustic characteristics.
\end{enumerate}
If any of the aforementioned conditions is violated, the processing is halted and an error message is displayed to the end user, instead of the actual report.

\subsection*{Utterance-level and Session-level Labeling}
Once the entire session is transcribed at the utterance level, we are able to employ text-based algorithms for the task of behavior code prediction. Both utterance-level and session-level behavior codes are predicted and provided back to the counselor as part of the feedback report, as described below.

\subsubsection*{Utterance-level Code Prediction}
We are focusing on counselor behaviors, so we only take into account the utterances assigned to the therapist according to SRR. Each one of those needs to be assigned a single code from the 9 target labels summarized in Table~\ref{table:misc_human_pc_map}. This is achieved through a BiLSTM network with attention mechanism~\citep{singla2018using} which only processes textual features. The input to the system is a sequence of word-level embeddings for each utterance. The recurrent layer exploits the sequential nature of language and produces hidden vectors which take into account the entire context of each word within the utterance. The attention layer can then learn to focus on salient words carrying valuable information for the task of code prediction, thus enhancing robustness and interpretability. The network was first trained on the TOPICS-CTT data using class weights to handle the problem of skewed code distribution in the data (Table~\ref{table:misc_human_pc_map}). The system was further fine-tuned by continuing training on the UCC$_{train}$ data in order to be suitably fitted to the UCC conditions.

\subsubsection*{Session-level Code Prediction}
Apart from the utterance-level codes, our system assigns a score to each one of the global codes of Table~\ref{table:globalMISC_descr}, ranging from 1 to 5. To that end, we represent the entire session, using the utterances assigned to both the therapist and the client, by the term frequency - inverse document frequency (tf-idf;~\citealp{salton1986intorduction}) transformation of unigrams, bigrams, and trigrams found within the session, excluding common stop words. Those features are $l_2$-normalized and passed to a Support Vector Regressor (SVR) which gives the final prediction. After hyper-parameter tuning, we chose polynomial SVR kernel (4th-degree) for acceptance and autonomy support, linear kernel for empathy, collaboration and evocation, and gaussian kernel for direction.

Contrary to the training approach followed for the utterance-level codes, here we train using only UCC data. The reason is that there is a discrepancy between the globals assigned by human raters to the TOPICS-CTT and the UCC sessions, since different coding procedures were followed. In particular, the TOPICS-CTT sessions were coded only across two global codes (empathy and MI spirit) following the Motivational Interviewing Treatment Integrity (MITI;~\citealp{moyers2016motivational}) coding scheme. Thus, due to the limited amount of training data (only 188 sample points---UCC sessions---in total), we apply a 5-fold cross validation scheme across the UCC dataset (from both coding trials) for any hyper-parameter tuning and we then keep those parameters to re-train using the entire UCC set.

\subsection*{Final Report}
After we have the automatically generated transcript and all the session-level and utterance-level predictions through our system, those are provided to the therapist as a feedback report through an interactive, web-based platform which we refer to as the Counselor Observer Ratings Expert for Motivational Interviewing (CORE-MI;~\citealp{hirsch2018s,imel2019design}). A video demonstration of the platform and its functionality is available at \url{www.youtube.com/watch?v=9fuvT9_azgw}.

CORE-MI features two main views, the session view and the report view~(Appendix~\ref{app:gui}). In the first one, the user can listen to the recording of the therapy, watch the video (if available) and read the generated transcript, which is scrollable and searchable. Additionally, they can keep notes linked to specific timestamps and utterances of the session. 

The report view provides the actual therapy session evaluation. The entire session timeline is presented in a form of a bar where talk turns of the two speakers are displayed in different colors. Hovering over a specific turn brings up the corresponding transcription and---in case the turn is assigned to the therapist---the corresponding MISC code(s). Based on the results reported later, we have decided to collapse the simple and complex reflections into one composite reflection (RE) label. The global behavior codes are also displayed, as well as a set of summary indicators which reflect the adherence to MI therapeutic skills. Those are the ratio of reflections (simple and complex) to questions (open and closed), the percentage of the open questions asked (among all the questions), the percentage of the complex reflections (among all the reflections), the percentage of each speaker's talking time, the MI adherence and the overall MI fidelity. MI adherence reflects the percentage of utterances where the counselor used MI-consistent techniques (e.g., asking open questions or giving advice with permission). Finally, the overall MI fidelity score is a composite metric rated on a 12-point scale that takes all the above into consideration and reflects the proficiency of the counselor to the different aspects of MI therapy. In particular, a provider can receive one point for passing pre-defined basic proficiency benchmarks and two points for passing advanced competency benchmarks across the following six measures of quality: empathy, MI spirit, reflection-to-question ratio, percentage of open questions, percentage of complex reflections, MI adherence. MI spirit is estimated as the average of evocation, collaboration, and autonomy support~\citep{houck2010misc}.

The main design characteristics of the CORE-MI platform have been tested in a past study~\citep{hirsch2018s,imel2019design} and results showed that the providers find the system easy to use and the feedback easy to understand. Additionally, most of the professional therapists that participated in the survey seemed excited about the potential opportunity to use such a system in clinical practice.

\section*{Results and Discussion}
\subsection*{Automatic Rich Transcription}
All the submodules of the transcription pipeline are evaluated on the two UCC test sets we have described (UCC$_{test_1}$, UCC$_{test_2}$), both individually and as part of the overall system. That way, we want to evaluate the performance of each one of the models, but, more importantly, investigate any error propagation that inevitably takes place.

\subsubsection*{VAD/Diarization}
During evaluation, VAD is usually viewed as part of a diarization system (e.g.,~\citealp{sell2018diarization}), so for evaluation purposes we consider our diarization model as the first component of the pipeline (frame-level VAD results are provided in Appendix~\ref{app:vad}). The standard evaluation metric for diarization is called Diarization Error Rate (DER;~\citealp{anguera2012speaker}) and it incorporates three sources of error: false alarms, missed speech, and speaker error. False alarm speech (the percentage of speech in the output but not in the ground truth) and missed speech (the percentage of speech in the ground truth but not in the output) are mostly associated with VAD. Speaker error is the percentage of speech assigned to the wrong speaker cluster after an optimal mapping between speaker clusters and true speaker labels. We estimate the DER on the UCC data using the \texttt{md-eval} tool which was developed as part of the Rich Transcription (RT) evaluation series (\url{www.nist.gov/itl/iad/mig/rich-transcription-evaluation}). We have used a forgiveness collar of $0.25\si{sec}$ around each speaker boundary, which is a standard practice~\citep{anguera2012speaker}, and the results are reported in Table~\ref{table:diar_res}.

\begin{table}[ht]
	\centering
	\caption{Diarization results ($\%$) for the test sets of the UCC data. {Diarization Error Rate (DER) is estimated as the sum of false alarm, missed speech, and speaker error rates.}}
	\label{table:diar_res}
	\begin{tabular}{ccccc}
		\toprule
		set & \spcellc{False\\Alarm} & \spcellc{Missed\\Speech} & \spcellc{Speaker\\Error} & DER \\
		\midrule
		$\text{UCC}_{test_1}$ & 13.7 & 0.4 & 6.9 & 21.0\\
		$\text{UCC}_{test_2}$ & 9.3 & 0.5 & 7.8 & 17.7\\
		\bottomrule
	\end{tabular}    
\end{table}

Even though the speaker confusion (speaker error rate) is on average low enough (lower than $8\%$), we should note that a per session analysis revealed that there are a few sessions where it is even higher than $45\%$. This means that diarization essentially failed for this handful of sessions, even though the human transcribers did not report any particular issues related, for example, to audio quality. 
	
Out of the three DER components, false alarm contributes most to the overall error, while the missed speech is minimal. Such a behavior is expected because of the specific implementation followed. In particular, we chose to concatenate together adjacent speech segments assigned to the same speaker, if there is not a silence gap between them greater that $1\si{sec}$. This step degrades the diarization result, since it labels short non-voiced segments as belonging to some speaker, thus introducing false alarms. However, it creates longer speaker-homogeneous segments, which is beneficial to ASR, and, hence, to the overall system, for two main reasons. First, it leads to more robust speaker adaptation since i-vectors are extracted from sufficiently long segments so that they capture a meaningful speaker representation. Second, it ensures larger language context, which means that the LM of the ASR system can choose the right word path with higher confidence.

\subsubsection*{Automatic Speech Recognition}
The evaluation of an ASR system is usually performed through the Word Error Rate (WER) metric which is the normalized Levenshtein distance between the ASR output and the ground truth transcript. This includes errors because of word substitutions, word deletions, and word insertions. For instance, word insertion rate is the number of new words included in the prediction which are not found in the original transcript over the total number of ground truth words. WER is calculated as the sum of those three error rates. Those errors are typically estimated for each utterance which is given to the ASR module and then summed up for all the evaluation data, in order to get an overall WER. However, when we analyze an entire therapy session which has been processed by the VAD and diarization sub-systems, the ``utterances'' are different than the ones identified by the human transcriber. In that case, we perform the evaluation at the session level, ignoring the speaker labels (from diarization) and concatenating all the utterances of the session. We do the same for the original transcript and we hence view the entire session as a ``single utterance'' for the purposes of ASR evaluation. The results are reported in Table~\ref{table:asr_res}. 

\begin{table}[ht]
	\centering
	\caption{{Automatic Speech Recognition (ASR)} results ($\%$) for the test sets of the UCC data: substitution, insertion, and deletion {rates, together with the total Word Error Rate (WER), estimated as the sum of those three.} Results are reported when using either the machine-generated segments (pipeline) or the ones derived by the manual transcriptions (oracle).}
	\label{table:asr_res}
	\begin{tabular}{cccccc}
		\toprule
		set & segmentation & subs & del & ins & WER \\
		\midrule
		\multirow{2}{*}{$\text{UCC}_{test_1}$} & oracle & 18.3 & 15.3 & 3.5 & 37.1 \\
		& pipeline & 20.0 & 13.9 & 4.3 & 38.1\\
		\midrule
		\multirow{2}{*}{$\text{UCC}_{test_2}$}& oracle & 14.3 & 13.8 & 2.3 & 30.4 \\
		& pipeline & 16.1 & 12.5 & 3.1 & 31.6 \\
		\bottomrule
	\end{tabular}    
\end{table}

As we can see, ASR performance is not severely degraded by any error propagation from the pre-processing step of diarization (WER is increased about $1\%$ absolute). Interestingly, even though the insertion rate is increased, the deletion rate is decreased when the machine-generated segments are provided. This is explained by the long segments constructed by the diarization algorithm and the post-processing of its output after concatenating consecutive segments. On the one hand, labeling silence or noise as ``speech'' associated with some speaker occasionally leads ASR to predict words where in reality there is no speech activity---thus increasing the insertion rate. On the other hand, this minimizes the probability of missing some words because of missed speech. Such deleted words may occur when providing the oracle segments because of inaccuracies during the construction of the ``ground truth'' through forced alignment.

We note that, even though the estimated error is high, WERs in the range reported ($30\%-40\%$) and even higher are typical in spontaneous medical conversations~\citep{kodish2018systematic}. Error analysis revealed that those numbers are inflated because of fillers (e.g. uh-huh, hmm) and other idiosyncrasies of conversational speech. It should be additionally highlighted that WER is a generic metric that gives equal importance to all the words, while for our end goal of behavior coding there are specific linguistic constructs which potentially carry more valuable information than others.

\subsubsection*{Speaker Role Recognition}
The described SRR algorithm operates at the session level, which means that, for evaluation purposes, it suffices to examine how many sessions are labeled correctly with respect to speaker roles. When oracle diarization information is provided, coupled either with the manual transcriptions or with the ASR results, our algorithm achieves a perfect recognition result for all the UCC$_{{test}_1}$ and UCC$_{{test}_2}$ sessions. When speaker segmentation and clustering is performed through the diarization algorithm of the processing system, the SRR module fails to find the right mapping between roles and speakers for seven sessions from the UCC$_{{test}_2}$ set. 

This behavior is associated with error propagation from the previous steps which is made apparent from the fact that the speaker error rate for those seven sessions is $42.5\%$ on average (std=$8.5\%$). Given the fact that we are dealing with dyadic conversational interactions, such a high speaker confusion essentially means that the diarization algorithm failed to sufficiently distinguish between the two speakers, probably because of similar acoustic characteristics. Thus, there is not enough reliable speaker-specific linguistic information that the SRR can use during the role assignment. This example of error propagation also highlights the need for quality assurance through specific safeguards at the early steps of the processing pipeline.

\subsubsection*{Utterance Segmentation}
The last step of the transcription pipeline is the utterance segmentation, which provides the basic units for behavioral coding. We get a rough indication of the quality of our segmentation process by estimating the correlation between the total number of utterances per session that have been assigned to the therapist by the human annotators and by the processing pipeline. The Spearman correlation between them, when all the UCC$_{test_1}$ and UCC$_{test_2}$ sessions are taken into account, is $0.478$ ($p<10^{-7}$). The number of the manually-defined utterances is usually higher than the number of the ones identified by our system, because the automated rich transcription module often fails to capture very short utterances (e.g., `yeah', `right', etc.). 

\subsection*{Quality Assurance}
According to the quality safeguards introduced, 16 out of the 112 sessions are flagged as ``problematic'' in our combined test set of UCC$_{test_1}$ and UCC$_{test_2}$. All of those do not meet the fourth condition, related to the minimum allowed speaking time attributed to each speaker. This means that in practice the processing would halt after the end of the diarization algorithm, with an error message displayed to the user. When we ran the entire set of 5,097 UCC recordings through the pipeline, 4,268 met all the four criteria and were successfully processed. 

It is interesting that, after excluding the ``problematic'' sessions from the test sets (UCC$_{test_1}$, UCC$_{test_2}$), the Spearman correlation between the total number of therapist utterances per session as assigned by the human coders vs. by the automated system is increased from 0.478 to 0.639. This is explained by the fact that, in several of those cases, poor diarization performance led the subsequent SRR module to assign almost the entire session to the client. As a result, the number of therapist-attributed utterances was much smaller than expected. 

\subsection*{Utterance-level and Session-level Labeling}
As in the case of the transcription pipeline submodules, we examine the effectiveness of the proposed models, both when provided with oracle information and when being part of the end-to-end system. In the following sections we discuss the results of the MISC code (utterance-level and session-level) prediction models.

\subsubsection*{Utterance-level Code Prediction}
When we use the manually transcribed data to perform utterance-level MISC code prediction, the overall $F_1$ score is $0.524$ for the UCC$_{test_1}$ and $0.514$ for the UCC$_{test_2}$ sets. The $F_1$ scores for each individual code are reported in Table~\ref{table:misc_res_f1}. As expected, the results are better for the highly frequent codes (Table~\ref{table:misc_human_pc_map}), such as FA, since the machine learning models have more training examples to learn from. On the other hand, the models do not perform as well for less frequent codes, such as MIN and MIA. However, comparing Table~\ref{table:misc_res_f1} and Table~\ref{table:interrater_alpha}, we can also see that for several of the codes that our system performs relatively poorly (e.g., RES, MIA, ST), the inter-annotator agreement is also considerably low. A notable example which does not follow this pattern is the non-adherent behavior (MIN) where our system achieves the lowest results among all the codes, while there is a substantial inter-annotator agreement ($\alpha=0.606$). This is partly because of the underrepresentation of the particular code (or cluster of codes) in the training and development sets. It may be also the case that pure linguistic information found in textual patterns may not be enough for the operationalization of the  particular code. This example suggests that a hybrid approach where machine learning methods are combined with knowledge-based rules from the coding manuals may be an interesting direction for future research. Finally, by examining the confusion matrices (not reported in this article), we realized that the system gets confused between the codes representing questions (QUC vs. QUO) and reflections (RES vs. REC), since those pairs of codes get usually assigned to utterances with several structural and semantic similarities.

\begin{table*}[ht]
	\centering
	\caption{$F_1$ scores for the predicted utterance-level codes using the manually transcribed UCC data.\\\hspace*{5cm}}\vspace*{-.4cm}
	\label{table:misc_res_f1}
	\begin{tabular}{cccccccccc}
		\toprule
		set & FA & GI & QUC & QUO & REC & RES & MIN & MIA & ST \\
		\midrule
		$\text{UCC}_{test_1}$ &  0.956 & 0.519 & 0.702 & 0.825 & 0.531 & 0.265 & 0.158 & 0.314 & 0.449 \\
		$\text{UCC}_{test_2}$&  0.951 & 0.462 & 0.588 & 0.786 & 0.465 & 0.186 & 0.273 & 0.439 & 0.474 \\
		\bottomrule
	\end{tabular}    
\end{table*}

\begin{figure*}[ht!]
	\centering
	\includegraphics[width=\textwidth]{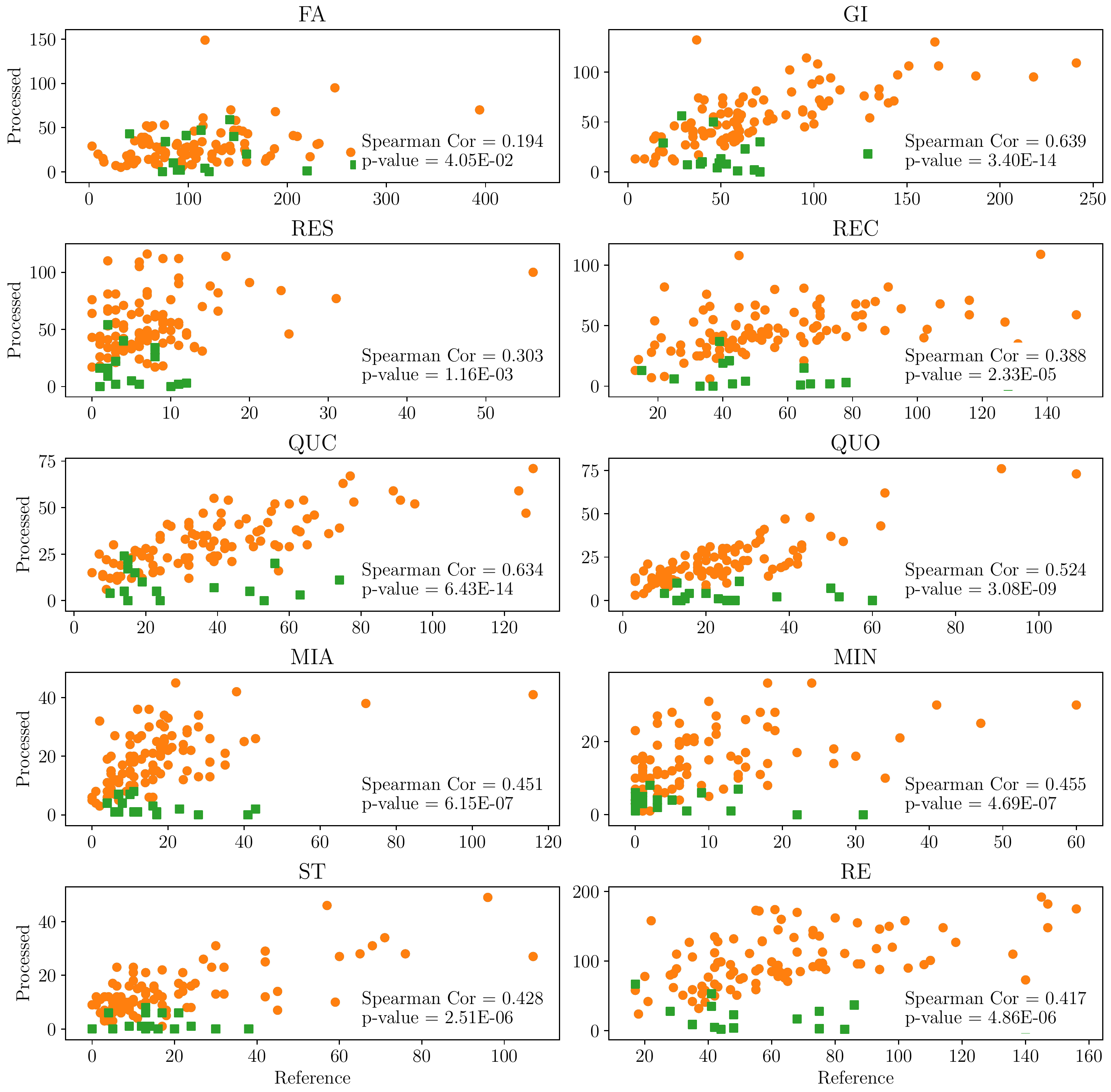}
	\caption{Count of each target MISC label per session when coded by humans (reference) and when processed by the pipeline. All the sessions in the UCC$_{test_1}$ and UCC$_{test_2}$ sets are shown and the correlation values are calculated based on all of them. The sessions flagged as problematic by the quality safeguards are denoted by square markers. RE is a composite label containing both RES and REC.}
	\label{fig:misc_freq}
\end{figure*}

The performance evaluation of the system when used within the pipeline is not straightforward, since the utterances given to the MISC predictor after the automatic transcription are not the same as the ones defined by the human transcribers. In that case, we use as a simple evaluation metric the correlation between the counts of each MISC label in the manual coding trial and in the automatically generated report. The results are illustrated in Figure~\ref{fig:misc_freq}. There is a statistically significant ($p<0.01$) positive correlation for all the codes, apart from FA. The Spearman correlation for the 9 codes is on average 0.446 (std=0.136), while if we don't take into consideration the sessions that did not meet the quality criteria, the correlation is increased to 0.566, on average (std=0.172).

The relatively low correlation and discrepancy in the counts between the manual and the automatically generated output for FA is striking, especially if we take into account the remarkably good results of the system when we do not use the entire pipeline (Table~\ref{table:misc_res_f1}). The reason is that FA is assigned to a lot of one-word utterances and talk turns. Our speech pipeline, however, often fails to capture turns of such short duration, which results in a smaller than expected frequency for the specific code. Another observed inconsistency is related to the code for simple reflections (RES), which seems to be assigned by our algorithm much more frequently than it actually occurs in the manually annotated data. As already mentioned, this is partly due to increased confusion between RES and REC. This becomes apparent if we merge all the reflections into one composite group (denoted as RE in Figure~\ref{fig:misc_freq}). 

\begin{figure*}[ht]
	\centering
	\includegraphics[width=\textwidth]{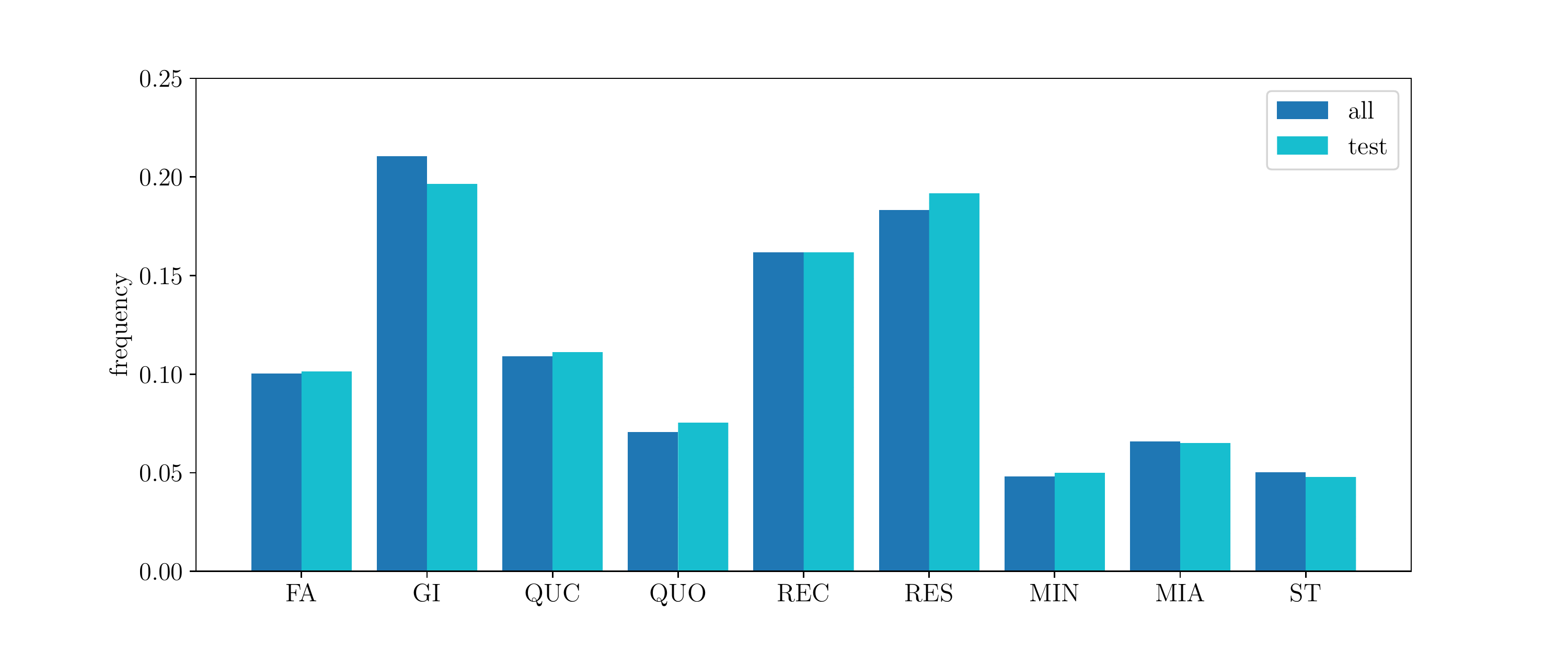}
	\caption{Frequency of the utterance-level MISC codes for all the UCC recordings processed and for the subset included in the UCC test sets. Only the sessions successfully processed (that met our quality criteria) are taken into consideration here. The total number of therapist-assigned utterances is about 1.2M for all the sessions (4,269 sessions) and 28K for only the sessions included in the UCC test sets (UCC$_{{test}_1}$ and UCC$_{{test}_2}$; 96 sessions).}
	\label{fig:misc_count}
\end{figure*}

The distribution of the MISC codes across all the 4,268 psychotherapy sessions that were successfully processed for this study is given in Figure~\ref{fig:misc_count}. As observed, the distribution is similar to the corresponding distribution if only the transcribed sessions included in the test sets (UCC$_{{test}_1}$ and UCC$_{{test}_2}$) are taken into consideration. This suggests that our test sets are indicative of the entire dataset and the evaluation analysis presented likely extends to previously unseen therapy sessions processed by the system.

\subsubsection*{Session-level Code Prediction}
As mentioned in the Materials and Methods section, the session-level code predictor is the only model where, due to the limited amount of training data, we apply a 5-fold cross validation scheme across the entire coded UCC dataset (all 188 sessions). The cross-validation results are reported in Table~\ref{table:globalMISC_res}. Results are given in terms of accuracy and averaged $F_1$ score, after the output of SVR is rounded to the closest integer in the range from 1 to 5 and after we collapse classes 1 and 2 together (due to the very limited number of sessions scored as 1 in the reference data). We also report the ‘within one’ accuracy, demonstrating whether the distance between the reference and predicted scores was at most one. In general, the predictive power of the models seems to be lower for the codes where the inter-rater reliability (Table~\ref{table:interrater_alpha}) is also low. Additionally, the performance is not severely affected by the usage of the speech pipeline, when compared to using the manual transcriptions.

\begin{table*}[ht]
	\centering
	\caption{Averaged $F_1$ scores and accuracy for the predicted session-level MISC codes using the manually transcribed (oracle) or the pipeline-generated data, based on a 5-fold cross validation scheme across all the UCC test data. The ‘within one’ accuracy demonstrates whether the distance between the predicted and reference scores was at most one point in the Likert scale.}
	\label{table:globalMISC_res}
	\begin{tabular}{l cc cc cc}
		\toprule
		metric & \multicolumn{2}{c}{$F_1$} & \multicolumn{2}{c}{acc} & \multicolumn{2}{c}{acc (`within 1')} \\
		\midrule
		ASR method & oracle & pipeline & oracle & pipeline & oracle & pipeline\\
		\midrule
		acceptance & 0.318 & 0.297 & 0.478 & 0.457 & 0.771 & 0.755\\
		empathy & 0.342 & 0.342 & 0.586 & 0.580 & 0.819 & 0.851\\
		direction & 0.380 & 0.261 & 0.426 & 0.389 & 0.740 & 0.697\\
		autonomy support & 0.303 & 0.261 & 0.495 & 0.451 & 0.878 & 0.840\\
		collaboration & 0.285 & 0.199 & 0.437 & 0.346 & 0.654 & 0.612\\
		evocation & 0.274 & 0.188 & 0.362 & 0.335 & 0.751 & 0.671\\
		\bottomrule
	\end{tabular}    
\end{table*}

\begin{figure*}[ht]
	\centering
	\includegraphics[width=\textwidth]{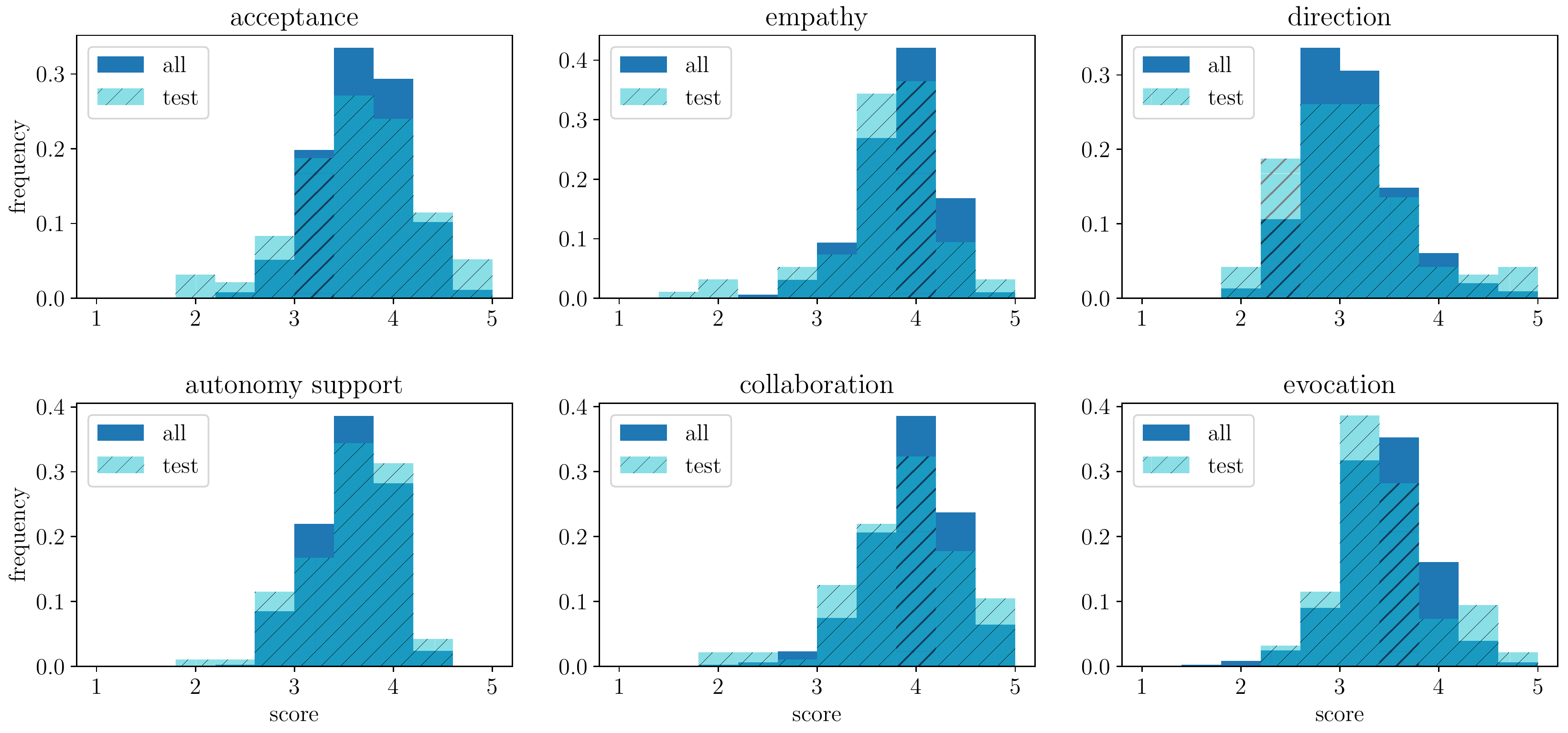}
	\caption{Distribution of the session-level MISC codes for all the UCC recordings processed and for the subset included in the UCC test sets. Only the sessions successfully processed (that met our quality criteria) are taken into consideration here.}
	\label{fig:globals_distr}
\end{figure*}

The distributions of the six global codes across all the 4,268 psychotherapy sessions that were successfully processed are given in Figure~\ref{fig:globals_distr}. All the codes, with the exception of direction, are skewed towards the higher scores of the scale (higher than 3). As was the case with the utterance-level codes (Figure~\ref{fig:misc_count}), we get a very similar distribution if we illustrate the results only for the sessions in the UCC test sets for which manual transcription and behavior coding information were available. This is indicative of the generalization of the system and its performance to future therapy sessions.

\section*{Limitations and Conclusions}
In this article we presented and analyzed a processing pipeline able to automatically evaluate recorded psychotherapy sessions. The application of such a system in real-world settings could guarantee the provision of fast and low-cost feedback. Performance-based feedback is an essential aspect both for training new therapists and for maintaining acquired skills, and can eventually lead to improved quality of services and more positive clinical outcomes. Additionally, being able to record, transcribe, and code interventions at large scale opens up ample opportunities for psychotherapy research studies with increased statistical power.

At the point of writing, we have processed a collection of more than 5,000 recordings, 4,268 of which met our quality criteria and are now accompanied by transcriptions and behavioral coding information. Both utterance-level and session-level MISC codes are available covering a wide range of behaviors (Figures~\ref{fig:misc_count} and~\ref{fig:globals_distr}). As we are planning on expanding our corpus with more data, we are confident that such a dataset will lead to novel interesting studies in the fields of psychotherapy, computational modeling, and their intersection. For example, the transcriptions of a subset of those data have been already used to study therapeutic alliance directly using text-based features~\citep{goldberg2020machine} or modeling clients and therapists as narrative characters~\citep{martinez2019identifying}. Even though we have here focused on Motivational Interviewing, the basic ideas of the speech processing pipeline remain the same for other dyadic interactions as well. For instance, the same modules analyzed in this article have been used to automatically transcribe and subsequently analyze cognitive behavior therapy sessions~\citep{chen2020feature}.

Despite the promising results presented here, we recognize that there is room for improvement in almost all the sub-modules of the pipeline. Our analysis showed that diarization failed for some of the sessions that human transcribers had no problem processing. Additionally, there was a consistent underrepresentation of verbal fillers (e.g., uh-huh) and the relevant MISC label (FA) in the automatically generated transcripts, as a result of the system struggling to capture and transcribe very short speaker turns. Moreover, the architecture design followed, where the various modules are trained independently and are then connected to form a pipeline, inevitably leads to error propagation. There are indications that alternative frameworks could reduce errors in specific cases, if for example diarization is aware of the different speaker roles~\citep{flemotomos2020linguistically} or if the two tasks of diarization and role recognition are performed simultaneously~\citep{flemotomos2018combined}.

For this work we only used text-based methods for behavioral coding. Acoustic features, however, and especially prosodic cues, play a major role in understanding language \citep{cutler1997prosody} and have been successfully used in the past for MISC code prediction~\citep{singla2018using,xiao2014modeling}. Recent studies have even shown that audio-only approaches, where word embeddings are directly learnt from spoken language, can yield improved results~\citep{singla2020towards}. Additionally, for the most part of our analysis, we have focused only on therapist characteristics. However, specific dialog attributes, such as language synchrony~\citep{lord2015more,nasir2019modeling} between the two involved parties (therapist vs. client)  and speech rate entrainment~\citep{xiao2015analyzing} can be proved useful for identifying therapy-related behaviors.

Another direction for potential future improvements is related to the modeling approach followed for the utterance-level codes. The system presented here treats all the codes evenly and employs a single neural architecture giving one output label for every utterance. However, since human coders often stack multiple codes for a single utterance (e.g., asking for permission to give advice [ADP] through a closed question [QUC]), a hierarchical algorithm which differentiates between codes with increasing granularity and allows for multiple codes per utterance may be useful. In such a scenario, a hybrid method which uses the modeling strength of neural networks and at the same time exploits knowledge-based information distilled from the coding manuals and clinical practice, can potentially improve the robustness and increase the interpretability of the results.This strategy would particularly benefit codes where our system performed relatively poorly (e.g., MIN, MIA; Table~\ref{table:misc_res_f1}), due to limited training examples or due to insufficient information captured just from available linguistic cues. Keeping in mind that psychotherapy is a dyadic interaction, incorporating contextual information from the client's neighboring utterances could also lead to performance improvements, especially for codes such as reflections (RES and REC) that depend semantically on client's language (Table~\ref{table:utteranceMISC_descr}).

Limitations imposed by the available number of training samples is a crucial aspect regarding any machine learning based model. Even though herein we present and use one of the largest available corpora constructed for the purpose of automatic behavioral coding, the performance of all the models involved is still critically dependent on the sample size. This is why we decided to use a lot of third-party sources, both for training the behavior code predictors and for any audio or language modeling needed for the transcription pipeline. Applying external datasets, however, was not possible for all the tasks. In particular, for the session-level code prediction, we only had the internal 188 labeled samples available and we, hence, decided to apply a cross-validation scheme with a statistical model (support vector regression) that does not require as much data as a more convoluted deep learning model to converge. In any case, all the results were reported on evaluation sets not seen during training, while the distributions of the predicted codes (Figures~\ref{fig:misc_count} and~\ref{fig:globals_distr}) suggest that those results are indicative of the performance on a much bigger dataset of therapy sessions.

An aspect of importance for our system is the quality assurance of the final evaluation report provided to the counselor. Being able to determine computational errors at an early stage and giving relevant warning messages to the user is an essential prerequisite before mental health practitioners trust computer-based tools and introduce them into clinical settings. We have already implemented several quality safeguards, with results indicating that they are towards the right direction. We are planning on implementing more confidence metrics, which take into account ASR and behavior coding results, apart from VAD and diarization. Human annotators can still be used for the sessions or parts of sessions for which confidence is low. Such manually-annotated sessions can be a valuable source of information to be used for further adapting our algorithms. That way, we can introduce an active learning scenario where the system incrementally becomes more accurate and reliable.

Likewise, it is important that we have evaluation metrics both for the individual modules and for the end-to-end system. Standard metrics, such as the Word Error Rate (WER) and the Diarization Error Rate (DER) used in ASR and in diarization, respectively, are useful during modeling in order to have benchmarks and quantifiable areas of improvement. However, they do not necessarily reflect the transcript quality from a user’s perspective~\citep{silovsky2012incorporation} and they are not always representative of the performance with respect to semantics and to clinical impact~\citep{miner2020assessing}. Qualitative surveys where experts share their opinions on the accuracy of the system output could assist highlighting specific areas of clinical importance on which the modeling efforts should focus.

We should here underline that our goal is to build a system that will not replace the human input, but will instead assist medical experts increasing efficiency and accuracy. Technology-based tools have seen a rapid rise in healthcare with applications ranging from safety surveillance and epidemiological data collection~\citep{cowie2017electronic} to clinical decision-making and treatment recommendations~\citep{sutton2020overview}. However, all those tools, and especially the ones focusing on conversational interactions, are not expected to replace care providers, but rather augment their capabilities~\citep{gangadharaiah2020conversational}. In the psychotherapy domain, an automatic evaluation platform, like the one we presented, would offer opportunities for ongoing self-assessment and self-improvement and would open new discussions on the development of specific skills between professionals or between trainees and supervisors. Additionally, even with a widespread usage of automatic psychotherapy evaluation systems, the community will still need skilled and objective behavioral coders, both for the evaluation and the training of the systems, since any machine learning algorithm is only as good as the training data we provide~\citep{caliskan2017semantics}.

In any case, it is essential that the users be adequately trained to understand the meaning of an automatically generated feedback and what the several scores represent. It has been reported that experienced counselors are more likely to be sceptical about the validity of their ratings~\citep{hirsch2018s}, as opposed to new and young therapists who may be attracted by the lure of machine learning, even without being fully aware of how their performance-based scores are estimated. Technology-based systems have the potential to transform mental healthcare. Being receptive to such a transformation should not mean uncritically accepting any machine-generated results. In fact, well-intentioned scepticism and criticism will accelerate the research in the field and will lead to an incremental improvement of the relevant technologies.

\section*{Open Practices Statement}
The original data collected for this study consist of real-world therapist-client sessions recorded at the University Counseling Center (UCC) of a large public Western university and have to remain within the UCC servers at all times for privacy reasons; thus they cannot be made publicly available. The psychotherapy data used from previous studies~\citep{tollison2008questions,baer2009agency,krupski2012testing,neighbors2012randomized,lee2013indicated,lee2014randomized} for adaptation are also protected and not publicly available. The speech corpora used to train the ASR system are either freely available or provided through the Language Data Consortium (LDC) to members and non-members for a fee (\url{www.ldc.upenn.edu}). In particular, Librispeech~\citep{panayotov2015librispeech} (\url{www.openslr.org/12}), TED-LIUM~\citep{rousseau2014enhancing} (\url{lium.univ-lemans.fr/ted-lium2}), and AMI~\citep{carletta2005ami} (\url{groups.inf.ed.ac.uk/ami/corpus}) are freely available to the community; Fisher English~\citep{cieri2004fisher} (Part 1: LDC2004S13 and Part 2: LDC2005S13), ICSI Meeting Speech~\citep{janin2003icsi} (LDC2004S02), WSJ~\citep{paul1992design} (Part 1: LDC93S6A and Part 2: LDC94S13A),  and 1997 HUB4~\citep{graff19971996} (LDC98S71) are provided through LDC. The Counseling and Psychotherapy Transcripts (without accompanying audio) that were used for some of the language-based modeling can be accessed on request at \url{alexanderstreet.com/products/counseling-and-psychotherapy-transcripts-series}.

Our models are trained on real-world, sensitive, and protected data. Thus, our trained models cannot be made publicly available. Acoustic feature extraction and acoustic modeling was performed using the Kaldi toolkit which is available at \url{github.com/kaldi-asr/kaldi}. The BeamformIt tool used for acoustic beamforming is available at \url{github.com/xanguera/BeamformIt}. Language models were built using the SRILM toolkit, available at \url{www.speech.sri.com/projects/srilm}. The neural network used for utterance-level code prediction was built on TensorFlow (\url{www.tensorflow.org}), while the tf-idf/SVR framework used for session-level code prediction made use of the \texttt{scikit-learn} Python library (\url{scikit-learn.org/stable}). The \texttt{md-eval} tool, developed by the National Institute of Standards and Technology (NIST) and used for diarization evaluation, is no longer available by NIST, but can be found at \url{github.com/nryant/dscore}. The estimation of Krippendorff's alpha ($\alpha$) for inter-rater reliability was based on the implementation available at \url{github.com/pln-fing-udelar/fast-krippendorff}.


\section*{Acknowledgements}
Funding was provided by the National Institutes of Health/National Institute on Alcohol Abuse and Alcoholism (R01 AA018673). 

\bibliography{ref}

\appendix

\section{System Training  Details}
\label{app:tech_detals}
The following sections provide some technical details related to the described system, including hyper-parameter values and training procedures. 

\subsection{Audio Feature Extraction}
MFCCs are extracted every $10\si{msec}$ using $25\si{msec}$-long windows.

\subsection{Voice Activity Detection}
The feed-forward neural network comprises two layers of 512 neurons each and sigmoid activation functions, before a final inference layer giving a frame-level probability. The input is a 13-dimensional MFCC vector characterizing a frame, spliced with a context of 30 neighboring frames (15+15). The frame-level VAD outputs are smoothed via a median filter of 31 taps. During this process, if the silence between any two contiguous voiced segments is less than 0.5sec, the corresponding segments are merged together. 

\subsection{Speaker Diarization}
Each voiced segment, as predicted by VAD, is partitioned uniformly into subsegments of length equal to $1.5\si{sec}$ with a shift of $0.25\si{sec}$. For each subsegment an x-vector is extracted using the pre-trained x-vector extractor found at \url{kaldi-asr.org/models/m6}. This was originally used to diarize telephone conversations and expects 23-dimensional MFCCs as input features. 
In particular, the first 5 layers of the neural architecture (x-vector extractor) operate at the frame level and are inspired from the Time-Delay Neural networks (TDNNs) where each layer sees a different temporal context. Those are followed by a statistics pooling layer that computes the mean and standard deviation vectors. Those are the inputs to a fully-connected layer that operates at the segment level before a final softmax inference layer that maps segments to speaker labels. The 128-dimensional embeddings used are the outputs of the final hidden layer and those are further mean- and length-normalized. 
The subsegments are finally clustered together according to a Hierarchical Agglomerative Clustering (HAC) approach with average linking, using PLDA as the similarity metric.
After this step, adjacent speech segments assigned to the same speaker are concatenated together into a single speaker turn, allowing a maximum of $1\si{sec}$ in-turn silence. 

\subsection{Automatic Speech Recognition}
The input feature vectors to the TDNN architecture are  40-dimensional MFCCs which are augmented by 100-dimensional i-vectors, extracted online through a sliding window.
First, word alignments are derived based on the GMM/HMM paradigm. 
The training data consists of the Fisher English, ICSI  Meeting  Speech, WSJ, 1997 HUB4, Librispeech, TED-LIUM, AMI, and TOPICS-CTT corpora. We use the officially recommended training subsets for Librispeech and TED-LIUM and the recommended training and development sets for AMI. We randomly choose $95\%$ of the available Fisher utterances and $80\%$ of the available ICSI, WSJ, and HUB4 utterances. We also use the 242 TOPICS-CTT$_{train}$ sessions described.in the paper. We have kept the rest of the combined dataset for internal validation and evaluation of the ASR system. Among the aforementioned corpora, TED-LIUM and the clean portion of Librispeech are augmented with speed perturbation, noise, and reverberation. The ASR AM was built and trained using the Kaldi speech recognition toolkit following the nnet3 `chain' setup. The two 3-gram LMs are trained using the SRILM toolkit and are interpolated with a mixing weight equal to 0.8 for the in-domain model and 0.2 for the background model.

\subsection{Speaker Role Recognition}
The required LMs are 3-gram models with Kneser-Ney smoothing, trained with the SRILM toolkit, using the TOPICS-CTT$_{train}$ and CPTS corpora with mixing parameters 0.8 and 0.2, respectively.

\subsection{Utterance-level Code Prediction}
The BiLSTM network was first trained on the TOPICS-CTT data using the Adam optimizer with a learning rate of 0.001 and an exponential decay of 0.9. The batch size was equal to 256 utterances. The system was trained on that dataset for 30 epochs with an early stopping strategy, keeping the model with the lowest validation loss. During the training process we used class weights inversely proportional to the class frequencies. The system was further fine-tuned by continuing training on the UCC$_{train}$ data. 

\section{CORE-MI Final Report Example}
\label{app:gui}

\begin{figure*}[ht]
	\centering
	\includegraphics[width=\textwidth]{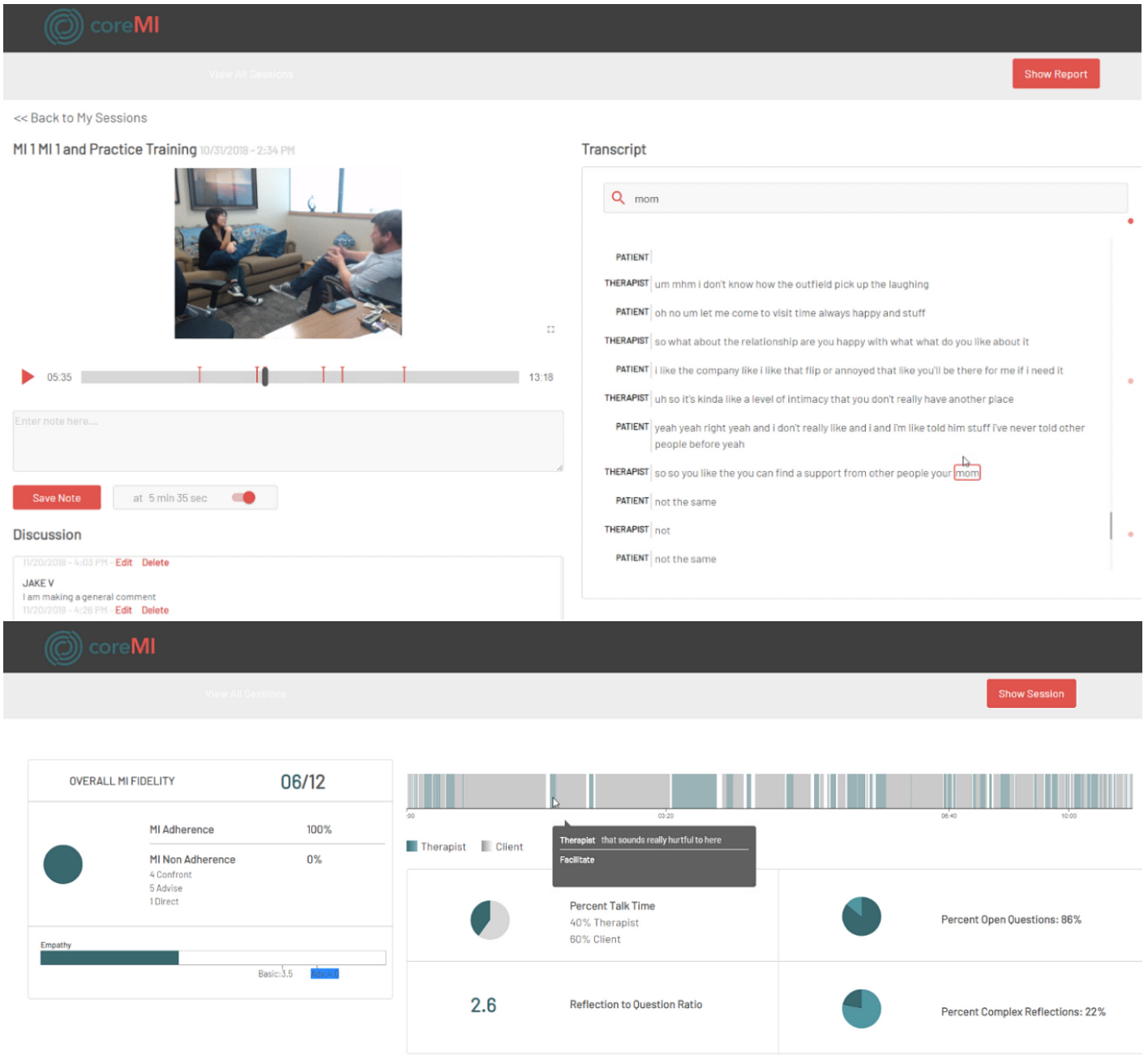}
	\caption{CORE-MI platform to provide therapy feedback. In the \emph{session view} (up) the user can listen to the recording, watch the video, read the generated transcription, and keep notes. In the \emph{report view} (down) MI fidelity scores are displayed.}
	\label{fig:corMI}
\end{figure*}

In Figure~\ref{fig:corMI}, the two main views of CORE-MI are displayed; the session view and the report view. In the first one, the user can listen to the recording of the therapy, watch the video (if available) and read the generated transcript, which is scrollable and searchable. The report view provides the actual therapy session evaluation. Among the session-level codes, only empathy is shown in the version displayed here.

\section{Voice Activity Detection - Frame Level Results}
\label{app:vad}

Voice Activity Detection (VAD) performance is typically incorporated in the evaluation of a diarization system in the form of false alarms and missed speech rates (Table~6 of the paper). However, especially in our system, those results are heavily influenced by post-processing steps and do not accurately represent VAD performance. The VAD results on the UCC data at the frame level are given in Table~\ref{table:vad_res}. In particular, the problem is treated as a binary classification one where each frame can take a binary value (voiced or unvoiced). Results are reported in terms of accuracy, precision and recall. The unweighted average recall is also reported and it is the metric used during hyper-parameter tuning.

\begin{table}[ht]
	\centering
	\caption{Voice Activity Detection (VAD) results ($\%$) at the frame level for the test sets of the UCC data. UAR is the Unweighted Average Recall and it was the main metric used for optimization of the VAD system.}
	\label{table:vad_res}
	\begin{tabular}{ccccc}
		\toprule
		set & accuracy & precision & recall & UAR \\
		\midrule
		$\text{UCC}_{test_1}$ & 85.4 & 83.0 & 92.6 & 83.6\\
		$\text{UCC}_{test_2}$ & 81.7 & 75.9 & 98.1 & 79.8\\
		\bottomrule
	\end{tabular}    
\end{table}

The reported precision of the system is partly affected by the ground truth construction. The VAD ground truth is obtained after aligning the audio and the transcripts, ignoring the speaker labels and allowing a maximum silence of $0.1\si{sec}$ between consecutive utterances. Since the unaligned words are ignored, and thus labeled as ``non-speech'', this results to an inflated count of false positives, i.e., frames which are (sometimes correctly) predicted as voiced, but are labeled as unvoiced by the ``ground truth''. 

\end{document}